\documentclass[12pt,letterpaper,makeidx]{article}
\usepackage[latin1]{inputenc}
\usepackage{amsmath}
\usepackage{amsfonts}
\usepackage{amssymb}
\usepackage{epsfig}
\usepackage{theorem}
\usepackage{oldgerm}
\usepackage{tabularx}
\usepackage{placeins}
\usepackage{amscd}
\usepackage{bm}
\usepackage{epic}
\usepackage{curves}
\usepackage{graphicx}
\usepackage{makeidx}
\usepackage{fancyhdr}
\usepackage{graphpap}
\usepackage{rotating}
\usepackage{bbm}
\usepackage{pstricks}
\usepackage[absolute]{textpos}
\usepackage{psfrag}
\usepackage{times}
\usepackage{dcolumn}
\usepackage{shadow}
\usepackage{multicol}
\usepackage[latin1]{inputenc}
\usepackage{oldgerm}
\fancypagestyle{plain}{%
\fancyhead{} 
}
\title{The CPT group of the spin-3/2 field}  
\author{B. Carballo Pérez\footnote{brendacp@nucleares.unam.mx}  and M. Socolovsky\footnote{socolovs@nucleares.unam.mx} \\ 
\small{ Instituto de Ciencias Nucleares, Universidad Nacional Autónoma de México,}\\
\small{ Circuito exterior, Ciudad Universitaria, 04510, México D.F., México}}
\date{}

\begin{document}
\maketitle
\textbf{Abstract}\\
\hspace{0.5cm}We find out that both the matrix and the operator CPT groups for the spin-3/2 field (with or without mass) are respectively isomorphic to $D_{4}\rtimes \mathbb{Z}_{2}$ and  $Q\times\mathbb{Z}_{2}$. These groups are exactly the same groups as for the Dirac field, though there is no a priori reason why they should coincide.\\

\newpage

\section{Introduction}
\hspace{0.5cm} The CPT group of the Dirac field in Minkowski space-time was obtained by Socolovsky in 2004 \cite{CPT group}. It were found two sets of consistent solutions for the matrices of charge conjugation ($C$), parity ($P$), and time reversal ($T$), which give the transformation of fields $\hat{\psi}_C(x)=C\hat{\bar{\psi}}^T(x)$, $\hat{\psi}_\Pi(x_\Pi)=P\hat{\psi}(x)$ and $\hat{\psi}_\tau (x_\tau)=T\hat{\psi}(x)^*$, where $x_\Pi=(t,-\bold{x})$ and $x_\tau=(-t,\bold{x})$. These sets are given by:
\begin{center}
a) $C_{D}=\pm\gamma^2\gamma^0$, $P_{D}=\pm i\gamma^0$, $T_{D}=\pm i\gamma^3\gamma^1$,\\
b) $C_{D}=\pm i\gamma^2\gamma^0$, $P_{D}=\pm i\gamma^0$, $T_{D}=\pm \gamma^3\gamma^1$.
\end{center}

Each of these sets generates a non abelian group of sixteen elements, respectively, $G_\theta^{(1)}\cong D_{4} \times \mathbb{Z}_{2}$ and $G_\theta ^{(2)} \cong 16E$,  where $D_{4}$ is the group of symmetries of the square and $16E$ is a non trivial extension of $D_{4}$ by $\mathbb{Z}_2$, isomorphic to a semidirect product of these groups. 

On the other hand, the quantum operators $\hat C$, $\hat P$ and $\hat T$, acting on the Hilbert space, generate a unique group $G_{\hat{\theta}} \cong Q\times \mathbb{Z}_2$, where $Q$ is the quaternion group.

With this in mind, we decided to find the $CPT$ group of the spin-3/2 field (Rarita-Schwinger field), for both massive and massless cases. This field could be useful for the description of compound objects (neglecting its structure in a first approximation), like the baryon decuplet components for spin-$3/2^{+}$ \cite{F.R.}, or for elementary fields such as the gravitino.

In order to describe 3/2-spin particles, the set of equations 
\begin{eqnarray}
(i \gamma^{\alpha} \partial_{\alpha}-m)\hat{\psi}^{\mu}(x)=0,\\
\gamma^{\mu}\hat{\psi}_{\mu}(x)=0,
\end{eqnarray}
where $\partial _\alpha={{\partial}\over {\partial {x_{\alpha}}}}$, is required. These equations are known as the Rarita-Schwinger equation \cite{R-Sch}. The first of these equation is a Dirac equation for each vector component of the vector spinor $\hat{\psi}^{\mu}$ and the second one is known as the subsidiary condition. Precisely due to the more complexity of these equations with respect to the Dirac field equation, there is not a priori any apparent reason why the CPT group for the Rarita-Schwinger field should coincide with those for the Dirac field.

\section{Parity}
\hspace{0.5cm} If we want to study the $P$ invariance of the Rarita-Schwinger equation, we need to repeat the analysis done for the Dirac equation in \cite{CPT group} and also consider the $P$ invariance of the subsidiary equation.

Multiplying this equation from the left by $P$, changing $\bold{x} \rightarrow -\bold{x}$ and inserting the unity, one obtains:

\begin{eqnarray}
P\gamma^{0}P^{-1}P\hat{\psi}_{0}(t,\bold{-x})+P\gamma^{1}P^{-1}P\hat{\psi}_{1}(t,\bold{-x})+P\gamma^{2}P^{-1}P\hat{\psi}_{2}(t,\bold{-x})\nonumber\\
+P\gamma^{3}P^{-1}P\hat{\psi}_{3}(t,\bold{-x})=0,
\label{subP}
\end{eqnarray}
but we need to take into account that the vector spinor changes by parity in the following way:

\begin{eqnarray}
\hat{\psi}_{\mu}(t,\bold{x})=\begin{pmatrix}
\hat{\psi}_0(t,\bold{x}) \cr \hat{\psi}_1(t,\bold{x}) \cr \hat{\psi}_2(t,\bold{x}) \cr \psi_3(t,\bold{x}) \cr
\end{pmatrix}\; \longrightarrow \; \hat{\psi}_{\mu \pi}(t,\bold{x})=\begin{pmatrix}
P\hat{\psi}_0(t,\bold{-x}) \cr -P\hat{\psi}_1(t,\bold{-x}) \cr -P\hat{\psi}_2(t,\bold{-x}) \cr -P\psi_3(t,\bold{-x}) \cr
\end{pmatrix};
\end{eqnarray}
where $\hat{\psi}_{\mu \pi}(t,\bold{x})$ can be written as $\hat{\psi}_{\mu \pi}(t,\bold{x})={\cal P}P\hat{\psi}_{\mu}(t,-\bold{x})$, with ${\cal P} \in O(1,3)$ given by:

\begin{equation}
{\cal P}=
\begin{pmatrix}
1 & 0 & 0 & 0\cr 0 & -1 & 0 & 0\cr 0 & 0 & -1 & 0\cr 0 & 0 & 0 & -1
\end{pmatrix}
\end{equation}
and $P\in D^{16}$, where $D^{16}$ is the Dirac algebra.

Substituting the components of $\hat{\psi}_{\mu \pi}(t,\bold{x})$ in (\ref{subP}) one obtains:

\begin{equation}
P\gamma^{0}P^{-1}\hat{\psi}_{0 \pi}(t,\bold{x})-P\gamma^{k}P^{-1}\hat{\psi}_{k \pi}(t,\bold{x})=0.
\end{equation}
This implies the constraints on $P$:

\begin{eqnarray}
P\gamma^{0}P^{-1}=\gamma^{0},\quad P\gamma^{k}P^{-1}=-\gamma^{k},
\label{Psub+}
\end{eqnarray}
or
\begin{eqnarray}
P\gamma^{0}P^{-1}=-\gamma^{0},\quad P\gamma^{k}P^{-1}=\gamma^{k}.
\label{Psub-}
\end{eqnarray}

The relations (\ref{Psub+}) are the same as those obtained from the Dirac equation, whose already known solution is $P_{D}=\pm i \gamma^{0}$, while from the relations (\ref{Psub-}) is obtained $P=P^{'}=z\gamma^{3}\gamma^{2}\gamma^{1}$, with $z\in \mathbb{C}^*=\mathbb{C}\setminus\{0\}$.

Following the same analysis as in \cite{CPT group}, there are two possibilities for each $P$:
\begin{center}
a) $P^2=+1 \ \Rightarrow \ z=\pm 1 \ \Rightarrow P^{'}=\pm \gamma^{3}\gamma^{2}\gamma^{1}$;\\
b) $P^2=-1 \ \Rightarrow \ z=\pm i \ \Rightarrow P^{'}=\pm i \gamma^{3}\gamma^{2}\gamma^{1}$.
\end{center}

In the first case: $P^{'\dagger}=P^{'}=P^{'-1}=-P^{'T} =-P^{'*}$ and in the second one: $P^{'\dagger}=-P^{'}=P^{'-1}=P^{'T} =-P^{'*}$.

\section{Charge conjugation}
\hspace{0.5cm} To study the $C$ invariance of the subsidiary equation, we must take the complex conjugate of this equation, multiply from the left by $C\gamma_{0}$ and insert the unit matrix. This is:

\begin{equation}
(C\gamma^{0})\gamma^{\mu *}(C\gamma^{0})^{-1}\hat{\psi}_{\mu C}(x)=0,
\end{equation}
where $\hat{\psi}_{\mu C}(x)=C\gamma^{0}\hat{\psi}_{\mu}^{*}(x)$.

In this case, the constraints on $C$ are:
\begin{eqnarray}
(C\gamma^{0})\gamma^{\mu *}(C\gamma^{0})^{-1}=\gamma^{\mu},
\label{Csub+}
\end{eqnarray}
or
\begin{eqnarray}
(C\gamma^{0})\gamma^{\mu *}(C\gamma^{0})^{-1}=-\gamma^{\mu}.
\label{Csub-}
\end{eqnarray}

Taking into account that $\gamma^0\gamma^{\mu*}\gamma^0=\gamma^{\mu T}$, one can find the solutions for the $C$ matrices. The equation with the negative sign is the same as in the Dirac case and leads to $C=C_{D}=\eta \gamma^{2}\gamma^{0}$; on the other hand, from equation (\ref{Csub+}) we arrive to:

\begin{equation}
C\gamma^{\mu T}C^{-1}=\gamma^{\mu},
\end{equation}
which leads to the solution $C=C^{'}=\eta \gamma^{3}\gamma^{1}$.

For $C^{'}$, a second application of the charge conjugation transformation is given by:
\begin{eqnarray}
(\hat{\psi}_C)_C &=&\hat{\psi}_{C^2}=C^{'}\hat{\bar{\psi}}_C^{T}=C^{'}(\hat{\psi}_C^{\dag}\gamma^{0})^T=C^{'}\gamma^{0}\hat{\psi}_C^*=C^{'}\gamma^0C^{'*}\gamma^{0}\hat{\psi}=-C^{'}C^{'*}\hat{\psi}\nonumber\\
&=&-\vert \eta \vert^2\gamma^{3}\gamma^{1}\gamma^{3*}\gamma^{1*}\hat{\psi}=\vert \eta \vert ^2(\gamma^3)^2(\gamma^{1})^2\hat{\psi}=\vert \eta \vert ^2\hat{\psi}.
\end{eqnarray}

Since the effect on $\hat{\psi}$ can be, at most, a multiplication by a phase, then $\eta\in U(1)$ and $C^{'}$ is unitary. This is:

\begin{equation}
C^{'}C^{'\dag}=\eta \gamma^3\gamma^{1}\bar{\eta}(\gamma^{3}\gamma^{1})^{\dagger}=\vert \eta \vert^2\gamma^3\gamma^{1}\gamma^1\gamma^{3}=\vert \eta \vert ^2 \gamma^3 (\gamma^1)^2 \gamma^3=\vert \eta \vert^2 1=1.
\end{equation}

Hence, for $C=C^{'}$ we also find that $\hat{\psi}_{C^2}=\hat{\psi}.$

As in \cite{CPT group}, due to a symmetry consideration between $\hat{\psi}_C=C^{'}\hat{\bar{\psi}}^T$ and $\hat{\bar{\psi}}_C =-\bar{\eta}^2 \hat{\psi}^T C^{'}$, it follows that $-\bar{\eta}^2=\pm 1$ and taking into account that $\vert \eta \vert ^4=1$, one obtains $\eta^2=\pm 1$, which implies that $\eta=\pm 1, \pm i$. 

Then, for the spin-3/2 field there are two possibilities for each $C$:
\begin{center}
a) $C^{'}=\pm \gamma^{3}\gamma^{1}, \qquad C_{D}=\pm \gamma^{2}\gamma^{0}$;\\
b) $C^{'}=\pm i\gamma^{3}\gamma^{1}, \qquad C_{D}=\pm i \gamma^{2}\gamma^{0}$.
\end{center}

\section{Time reversal}
\hspace{0.5cm} We start again from the subsidiary equation, change $t\rightarrow -t$ and take the complex conjugate:
\begin{eqnarray}
T\gamma^{0*}T^{-1}T\hat{\psi}_{0}(-t,\bold{x})^{*}+T\gamma^{1*}T^{-1}T\hat{\psi}_{1}(-t,\bold{x})^{*}+T\gamma^{2*}T^{-1}T\hat{\psi}_{2}(-t,\bold{x})^{*}\nonumber\\
+T\gamma^{3*}T^{-1}T\hat{\psi}_{3}(-t,\bold{x})^{*}=0,
\label{subT}
\end{eqnarray}
but we need to take into account that the vector spinor changes by time inversion in the following way:

\begin{eqnarray}
\hat{\psi}_{\mu}(t,\bold{x})=\begin{pmatrix}
\hat{\psi}_0(t,\bold{x}) \cr \hat{\psi}_1(t,\bold{x}) \cr \hat{\psi}_2(t,\bold{x}) \cr \psi_3(t,\bold{x}) \cr
\end{pmatrix}\; \longrightarrow \; \hat{\psi}_{\mu \tau}(t,\bold{x})=\begin{pmatrix}
-T\hat{\psi}_0(-t,\bold{x})^{*} \cr T\hat{\psi}_1(-t,\bold{x})^{*} \cr T\hat{\psi}_2(-t,\bold{x})^{*} \cr T\psi_3(-t,\bold{x})^{*} \cr
\end{pmatrix};
\end{eqnarray}
where $\hat{\psi}_{\mu \tau}(t,\bold{x})$ can be written as  $\hat{\psi}_{\mu \tau}(t,\bold{x})={\cal T}T\hat{\psi}_{\mu}(-t,\bold{x})^{*}$, with ${\cal T} \in O(1,3)$ given by:

\begin{equation}
{\cal T}=
\begin{pmatrix}
-1 & 0 & 0 & 0\cr 0 & 1 & 0 & 0\cr 0 & 0 & 1 & 0\cr 0 & 0 & 0 & 1
\end{pmatrix}
\end{equation}
and $T\in D^{16}$.

Substituting the components of $\hat{\psi}_{\mu \tau}(t,\bold{x})$ in (\ref{subT}) one obtains:

\begin{equation}
-T\gamma^{0*}T^{-1}\hat{\psi}_{0 \tau}(t,\bold{x})+T\gamma^{k*}T^{-1}\hat{\psi}_{k \tau}(t,\bold{x})=0,
\end{equation}
from which we can deduce the constraints on $T$:

\begin{eqnarray}
T\gamma^{0}T^{-1}=\gamma^{0},\quad T\gamma^{k*}T^{-1}=-\gamma^{k},
\label{Tsub+}
\end{eqnarray}
or
\begin{eqnarray}
T\gamma^{0}T^{-1}=-\gamma^{0},\quad T\gamma^{k*}T^{-1}=\gamma^{k}.
\label{Tsub-}
\end{eqnarray}

The relations (\ref{Tsub+}) are the same as those obtained from the Dirac equation, whose already known solution is $T_{D}=e^{i\lambda}\gamma^{3}\gamma^{1}$, while from the relations (\ref{Tsub-}) is obtained $T=T^{'}=w\gamma^{2}\gamma^{0}$, with $w\in \mathbb{C}^*=\mathbb{C}\setminus\{0\}$.

For $T=T^{'}$, applying $\tau$ twice, we arrive to:
\begin{equation}
\hat{\psi}(t,\bold{x})\to \hat{\psi}_\tau(t,\bold{x})=T^{'}\hat{\psi}(-t,\bold{x})^*\to T^{'}(T^{'}\hat{\psi}(t,\bold{x})^*)^*=T^{'}T^{'*}\hat{\psi}(t,\bold{x}).
\end{equation}
and taking into account that
\begin{equation}
T^{'}T^{'*}=\vert w \vert ^{2}\gamma^{2}\gamma^{0}(\gamma^{2}\gamma^{0})^{*}=-\vert w \vert ^{2}\gamma^{2}\gamma^{0}\gamma^{2}\gamma^{0}=\vert w \vert ^{2}\gamma^{2}\gamma^{2}\gamma^{0}\gamma^{0}=-\vert w \vert ^{2} 1,
\end{equation}
it follows that $\hat{\psi}_{\tau^2}=-\hat{\psi}$, by a similar argument to that used for $C$. 

Thus, $T^{'}T^{'*}=-1$, which implies $T^{'*}=-T^{'-1}$ and $w \in U(1)$. Then, $T^{'}=e^{i\lambda}\gamma^{2}\gamma^{0}$ y $T^{'\dagger}=e^{-i\lambda}\gamma^{2}\gamma^{0}$.

\section{Matrix and operator CPT groups}
\hspace{0.5cm}In summary, one has both the two sets of matrices:

\begin{center}
a) $C_{D}=\pm\gamma^2\gamma^0, \qquad P_{D}=\pm i\gamma^0, \qquad T_{D}=\pm i\gamma^3\gamma^1$,\\
b) $C_{D}=\pm i\gamma^2\gamma^0, \qquad P_{D}=\pm i\gamma^0, \qquad T_{D}=\pm \gamma^3\gamma^1$;
\end{center}
and the set: $C^{'}, P^{'}, T^{'}$, satisfying the subsidiary equation. But, only the matrices $C_{D}, P_{D}, T_{D}$, also satisfy the Dirac type equation. That is why they are the matrices which conform the matrix CPT group of the spin-3/2 field with mass.

If we take the zero mass limit of the Dirac type equation,
\begin{equation}
i\gamma^{\alpha}\partial_{\alpha}\hat{\psi}(x)=0,
\label{ecDirac}
\end{equation}
and analyze its behavior under parity, charge conjugation and time reversal, as we did for the subsidiary condition, we have, respectively:

\begin{equation}
i(P\gamma^0 P^{-1}\partial_0-P\gamma^{i}P^{-1} \partial_i)\hat{\psi}_\Pi(x)=0,
\end{equation}

\begin{equation}
(i\partial_\mu+q A_\mu)(C\gamma^0)\gamma^{\mu*}(C\gamma^0)^{-1}\hat{\psi}_C(x)=0.
\end{equation}

\begin{equation}
i(\gamma^{0*}{{\partial}\over {\partial t}}-\gamma^{k*}{{\partial}\over{\partial x^k}})\hat{\psi}_\tau(x)=0.
\end{equation}

From the above equations we can then obtain the corresponding restrictions on the $C$, $P$ and $T$ matrices,  respectively:

\begin{eqnarray}
P\gamma^{0}P^{-1}=\pm\gamma^{0},\quad P\gamma^{k}P^{-1}=\mp\gamma^{k},\nonumber\\
C\gamma^{\mu T}C^{-1}=\pm\gamma^{\mu},\nonumber\\
T\gamma^{0}T^{-1}=\pm\gamma^{0},\quad T\gamma^{k*}T^{-1}=\mp\gamma^{k}.
\end{eqnarray}

These relations generate the same sets of matrices $C_{D}, P_{D}, T_{D}$ and
$C^{'}, P^{'}, T^{'}$ which gave the subsidiary condition. But if we take into account that $\hat{\bar{\psi}}\hat{\psi}$ is the charge density operator, for $C=C^{'}$, it follows that:
\begin{eqnarray}
(\hat{\bar{\psi}}\hat{\psi})_C=\hat{\bar{\psi}}_C\hat{\psi}_C=-\bar{\eta}^2\hat{\psi}^T C^{'}C^{'}\hat{\bar{\psi}}^T=-\bar{\eta}^2C^{'2}(\hat{\bar{\psi}}\hat{\psi})^T\nonumber\\
=-\bar{\eta}^2C^{'2}\hat{\bar{\psi}}\hat{\psi}
=(\bar{\eta}\eta)^2\hat{\bar{\psi}}\hat{\psi}=\vert \eta \vert ^4\hat{\bar{\psi}}\hat{\psi}=-\hat{\bar{\psi}}\hat{\psi};
\end{eqnarray}
from which $\vert \eta \vert ^4=-1$, which is a contradiction.  Hence, the matrix $C=C^{'}$ must be discarted.

It was demostrated in \cite{CPT group} that $C$ and $P$ must fulfill the relation:
\begin{equation}
C(P^{-1})^T C^{-1}=P,
\end{equation}
while $C$ and $T$ are related by:
\begin{equation}
CT^*=TC^*.
\end{equation}

Due to the fact that each component of the vector spinor satisfies a Dirac type equation, the above relations also hold in our case. That is why the set of matrices: $C^{'}, P^{'}, T^{'}$ are also discarded for the massless case.

In order to find the operator CPT group for the spin-3/2 field (with or without mass), we follow the same procedure developed in \cite{CPT group}, for the case of the corresponding CPT group of the Dirac field.

Taking $\hat A$ and $\hat B$ as any of the operators $\hat C_{D}$, $\hat P_{D}$ and $\hat T_{D}$; and $\hat{\psi}$, as each component of the vector spinor $\hat{\psi}^{\mu}(x)$, the relations:
\begin{equation}
\hat A \cdot \hat{\psi}=\hat A^{\dag} \hat{\psi} \hat A
\end{equation}
and
\begin{equation}
(\hat A *\hat B)\cdot \hat{\psi}=(\hat A \hat B)^{\dag}\hat{\psi} (\hat A \hat B),
\end{equation}
can be defined.

Using the above expressions and with support in the matrix CPT group, through the formulas that link matrices with operators:
\begin{eqnarray}
P\hat{\psi}(t,-\bold{x})={\hat P}^{\dag}\hat{\psi}(t,\bold{x}){\hat P}, \nonumber\\
C\hat{\bar{\psi}}^T(x)={\hat C}^{\dag}\hat{\psi}(x){\hat C}, \nonumber\\
T\hat{\psi}(-t,\bold{x})^*={\hat T}^{\dag}\hat{\psi}(t,\bold{x})^{\dag}{\hat T};
\end{eqnarray}
the relations:
\begin{eqnarray}
\hat{P_{D}}*\hat{P_{D}}=-1, \quad \hat{C_{D}}*\hat{C_{D}}=1, \quad \hat{T_{D}}*\hat{T_{D}}=-1, \nonumber\\
\hat{T_{D}}*\hat{P_{D}}=-\hat{P_{D}}*\hat{T_{D}},  \; \hat{C_{D}}*\hat{P_{D}}=\hat{P_{D}}*\hat{C_{D}},  \; \hat{C_{D}}*\hat{T_{D}}=\hat{T_{D}}*\hat{C_{D}},
\label{relopC}
\end{eqnarray}
were obtained in \cite{CPT group}; from which it is possible to build, also using the property of associativity, the multiplication table for the operator CPT group.

It was also demostrated in \cite{CPT group} that only the second of the two solutions for the matrix group ($G_\theta ^{(2)} \cong 16E \cong D_{4}\rtimes \mathbb{Z}_{2}$ ), is compatible with the operator group ($G_{\hat{\theta}} \cong Q\times \mathbb{Z}_2$).

In summary, we showed that both the matrix and the operator CPT groups for the spin-3/2 field (with or without mass) coincide with the corresponding groups for the Dirac field.

\section{Acknowledgment}
\hspace{0.5cm} This work was partially support by the project PAPIIT IN 118609-2, DGAPA-UNAM, M\'exico. B. Carballo P\'erez also acknowledge financial support from CONACyT, M\'exico. The authors thank Prof. O. S. Zandrón from IFIR, Rosario, Argentina, for suggesting this work.


\end{document}